\documentclass[preprint,12pt,authoryear]{elsarticle}

\usepackage{amssymb}
\usepackage{amsmath}

\usepackage{graphicx}%
\usepackage{multirow}%
\usepackage{amsmath,amssymb,amsfonts}%
\usepackage{amsthm}%
\usepackage{mathrsfs}%
\usepackage[title]{appendix}%
\usepackage{xcolor}%
\usepackage{textcomp}%
\usepackage{manyfoot}%
\usepackage{booktabs}%
\usepackage{algorithm}%
\usepackage{algorithmicx}%
\usepackage{algpseudocode}%
\usepackage{listings}%
\usepackage{siunitx}
\usepackage{hyperref}
\usepackage{cleveref}

\newcommand{\au}{\,\mathrm{au}}
\newcommand{\km}{\,\mathrm{km}}
\newcommand{\meter}{\,\mathrm{m}}
\newcommand{\cm}{\,\mathrm{cm}}
\newcommand{\mm}{\,\mathrm{mm}}

\newcommand{\second}{\,\mathrm{s}}

\newcommand{\kg}{\,\mathrm{kg}}
\newcommand{\g}{\,\mathrm{g}}

\journal{Icarus}

\begin{document}

\begin{frontmatter}

\title{Instantaneous thermally-driven erosion can explain dearth of dark near-Sun asteroids} 

\author[label1]{Georgios Tsirvouils} 
\author[label1,label2]{Mikael Granvik} 
\author[label1]{Leonard Schirner} 
\author[label1]{Athanasia Toliou} 
\author[label1]{Jooyeon Geem} 
\author[label3]{Axel Hagermann} 

\affiliation[label1]{organization={Asteroid Engineering Laboratory, Luleå University of Technology},
            addressline={Bengt Hultqvists vag 1}, 
            city={Kiruna},
            postcode={98192}, 
            state={},
            country={Sweden}}

\affiliation[label2]{organization={Department of Physics, University of Helsinki},
            addressline={P.O. Box 64}, 
            city={Helsinki},
            postcode={00014}, 
            state={},
            country={Finland}}

\affiliation[label3]{organization={Luleå University of Technology},
            addressline={Bengt Hultqvists vag 1}, 
            city={Kiruna},
            postcode={98192}, 
            state={},
            country={Sweden}}

\begin{abstract}
Recent models of the near-Earth asteroid population show that asteroids must be super-catastrophically destroyed when they evolve to orbits with perihelion passages well inside of Mercury's orbit. The heliocentric distances at which the disruptions typically occur are tens of solar radii, which is too far from the Sun for asteroids to be destroyed by sublimation and tidal disruption. The typical disruption distance also appears to be larger for darker asteroids. Here, by carrying out irradiance experiments in vacuum that replicate the conditions in the near-Sun environment, we show that CI meteorite simulants are destroyed within minutes when exposed to the level of solar irradiance encountered at heliocentric distances of up to about 0.2~au. Our results provide an explanation for the scarcity of dark, carbonaceous asteroids with perihelion distances less than 0.2~au, and for the observed mass-loss rate of the asteroid-like object 322P/SOHO~1 assuming its composition is similar to CI carbonaceous chondrites.
\end{abstract}


\end{frontmatter}




\section{Introduction}\label{intro}

All asteroids that reach sufficiently small perihelion distances $q$ disrupt super-catastrophically, i.e. they disintegrate into millimeter-sized fragments \citep{Granvik2016,Wiegert2020}. The critical perihelion distance at which the disruption occurs depends on the size of the asteroid and its material composition, and is about $q\lesssim0.2\au\ (\approx\text{43 solar radii})$ for decameter-sized and larger asteroids \citep{Granvik2016,Nesvorny2024}. In the following, we call these objects 'near-Sun asteroids' or NSAs. NSAs with low geometric albedos, that is, those that resemble carbonaceous chondrites, tend to be destroyed at larger perihelion distances \citep{Granvik2016}. The almost complete lack of NSAs with perihelion distances smaller than the critical value, including those with low geometric albedos, cannot be explained as an observational bias \citep{Granvik2016}.

Understanding the process that destroys NSA is important, because it affects orbit and size distributions of NEOs in general \citep{Granvik2016,Nesvorny2024}, and it is a source for meteoroids on Earth-crossing orbits \citep{Wiegert2020} that may pose a hazard to spacecraft. Understanding the process of disruption may allow for constraining the bulk composition of NSAs---and asteroids in general since NSAs are sourced from the asteroid belt---because the susceptibility for destruction appears to be correlated with composition \citep{Granvik2016}.

The exact disruption mechanism(s) are still unknown, but some proposed hypotheses exist. These explanations can be divided into two broad categories: impact-driven disruption \citep{Wiegert2020}, and thermally-driven disruption \citep{Granvik2016,Molaro2020}. Whilst the capability of meteoroid impacts to erode the surfaces of asteroids is well established in terms of theoretical and numerical models as well as impact experiments \citep{Szalay2018}, there has been limited knowledge about the spatial distribution of meteoroids near the Sun. The Parker Solar Probe is now providing the first direct measurements of meteoroid number densities at distances in the range of about $0.05-1\au$ from the Sun, which are relevant for testing the impact explanation \citep{Shen2024}.

Compared to impact erosion, the theory and models of thermally-driven destruction of NSAs are less evolved, because its potential importance has only recently been understood. However, there are examples of near-Sun asteroids showing signs of activity that may be thermally driven, and that could lead to the destruction of entire asteroids over time. 

The prime example of an active NSA is (3200)~Phaethon, which shows recurrent activity for a few days during its perihelion passage at $q\sim0.14\au$ \citep{Li2013,Jewitt2013}. Phaethon is considered the parent body of the meteoroid stream that produces the Geminid meteor shower \citep{Williams1993}, although it has been shown that Phaethon's current level of activity is not enough to sustain the Geminid meteoroid stream \citep{Jewitt2019}. Another example is 322P/SOHO~1 (P/1999~R1) \citep{hoenig2006}, for which the brightness increase during its perihelion passage at a heliocentric distance of $0.05\au$ is similar to that observed for comets, and it hence received a cometary designation. However, SOHO~1 does not have a visible coma or tail during the perihelion passage in SOHO/LASCO images, and activity was not detected when the 150--320-meter-diameter object was observed with the Lowell Discovery Telescope (then Discovery Channel Telescope), ESO's Very Large Telescope, and the Spitzer Space Telescope at heliocentric distances ranging from $1.2\au$ to $2.2\au$, that is, distances were comets usually show significant activity due to the sublimation of water ice \citep{knight2016}. The $V-R$ and $R-I$ colors suggest a V-type or Q-type asteroid \citep{knight2016}, but measurement uncertainties do not allow for a definite classification, and a C-type classification could still be possible. Whereas few-hundred-meter-diameter asteroids are expected to disrupt before evolving to $q\sim0.05\au$ \citep{Granvik2016,Nesvorny2024}, 322P could have originally been a larger body that can survive the evolution to a smaller perihelion distance than its current size suggests.

A major obstacle for obtaining a better understanding of the disruptive effect(s) that solar irradiance has on NSAs is that very few experimental results have been published on the subject. For example, there are no published experimental results on the rate of disruption of asteroid-like materials as a function of temperature, or, equivalently, distance from the Sun. The Space and High-Irradiance Near-Sun Simulator \citep[SHINeS;][]{Tsirvoulis2022} is a facility set up to study the behavior of natural materials in close proximity of the Sun. Its main components are a vacuum chamber, a solar simulator capable of achieving a maximum irradiance of $\sim35\ W/cm^2$ and measuring equipment such as video cameras, a residual gas analyser and temperature sensors. Whereas previous experimental works have focused on, e.g., the evolution of reflectance spectra with increasing temperature \citep{Sidhu2023}, as well as modeling and measuring the outgassing of different species \citep{masiero2021volatility}, we carried out SHINeS experiments with a focus on the mechanical and temporal aspects of the destruction of simulants of carbonaceous chondrites of the Ivuna type \citep[CI;][]{Britt2019} to understand if thermally-driven processes can explain the dearth of dark asteroids on near-Sun orbits. A focus point for our experiments was the perihelion distance of Phaethon, but we also carried out experiments at shorter and greater distances from the Sun to understand how the destruction rate varies with heliocentric distance.

\section{Methods}

\subsection{Sample preparation}

The CI simulant used was developed by the Exolith Lab at the University of Central Florida \citep{Britt2019}. This simulant scored high figures of merit across a wide range of properties when compared to the meteorite its development was based on, Orgueil, confirming its high fidelity \citep{Metzger}. Table~\ref{tab:min} presents the mineralogy of the simulant as well as its bulk chemistry as measured by X-ray fluorescence. To ensure repeatable experiments the simulant was prepared according to the default recipe provided by the manufacturer: four mass units of the simulant material were mixed by hand with one mass unit of de-ionized water, forming a thick paste which was then spread evenly in a 10-mm thick layer onto aluminum baking trays. The trays were then transferred into an oven at \SI{60}{\degreeCelsius} to cure for 48~hours. The material was then cut into pieces to obtain identical protopellets with dimensions $5\mm \times 20\mm \times 20\mm$. The horizontal dimensions of the sample pellets are dictated by the width of the light beam and its spatial uniformity \citep{Tsirvoulis2022}. The protopellets were then brought to a thickness of $5\mm$ by sanding both the top and bottom sides on a flat abrasive surface. The process ensured that all the sample pellets originate from the core of the "cake", in an effort to ensure the highest compositional and mechanical uniformity possible.

\begin{table}[ht]
\centering

\caption{CI Simulant mineralogy (left) and bulk chemistry (right), adapted from Tables 2 and 3 of \citet{Britt2019}.}
\begin{tabular}{|lr@{\hspace{0.5cm}}|@{\hspace{0.5cm}}lr|}
\hline
Mineral & CI Simulant (wt\%) & Oxide & CI Simulant (wt\%) \\
\hline
Mg-serpentine & 48.0 & SiO$_2$ & 25.0 \\
Epsomite & 6.0 & TiO$_2$ & 0.5 \\
Magnetite & 13.5 & Al$_2$O$_3$ & 3.1 \\
Palygorskite & 5.0 & FeO$_T$ & 25.8 \\
Olivine & 7.0 & MnO & 0.3 \\
Pyrite & 6.5 & MgO & 30.2 \\
Vermiculite & 9.0 & CaO & 3.0 \\
Coal & 5.0 & Na$_2$O & 6.4 \\
 & & K$_2$O & 0.4 \\
 & & P$_2$O$_5$ & 0.4 \\
 & & Cr$_2$O$_3$ & 0.2 \\
 & & SO$_3$ & 4.9 \\
\hline
 Total & 100.00 & Total & 100.0 \\
\hline
\end{tabular}
\label{tab:min}
\end{table}

We carried out a test to assess whether the described procedure introduces any anisotropic behavior to the simulant: we prepared and baked a thicker layer of the paste, from which we cut three sample pellets of the same dimensions as before, but in a vertical direction. We could not observe any differences in the experiment outcomes compared to the regular horizontal samples. This outcome leads us to conclude that our preparation procedure did not introduce any significant anisotropic properties to the sample protopellets.

\subsection{Experiment procedure}

The configuration of the light source and the lenses was first adjusted to produce an irradiance corresponding to a heliocentric distance of 0.14~au. The irradiance was measured using a thermal power detector with a circular aperture of diameter $d=12.5\mm$ at the center of the light spot at the focal plane, where the sample is located and the uniformity was then measured using brightness mapping \citep{Tsirvoulis2022} to be within 95\% of the nominal value for an area approximately $30\mm \times 30\mm $ which was larger than the sample itself. We followed the same process when the optical path configuration was altered to simulate a different heliocentric distance.

Each sample pellet was next loaded onto an insulating slab of aluminum oxide, and gently secured in place with two aluminum holders to restrict potential movement during the experiment. The whole assembly was placed into the vacuum chamber, which was then evacuated to a target pressure of $P_t<5\cdot10^{-5}\,$mbar.  When the vacuum chamber reached the target pressure, the sample was insolated at the specified irradiance, and the entire process recorded with two high-speed cameras. At each simulated heliocentric distance we performed 3--4 identical experiments. 

\section{Results}

\subsection{Qualitative analysis of the destruction process at a heliocentric distance of 0.14au}

\begin{figure}
\centering
\includegraphics[width=\textwidth]{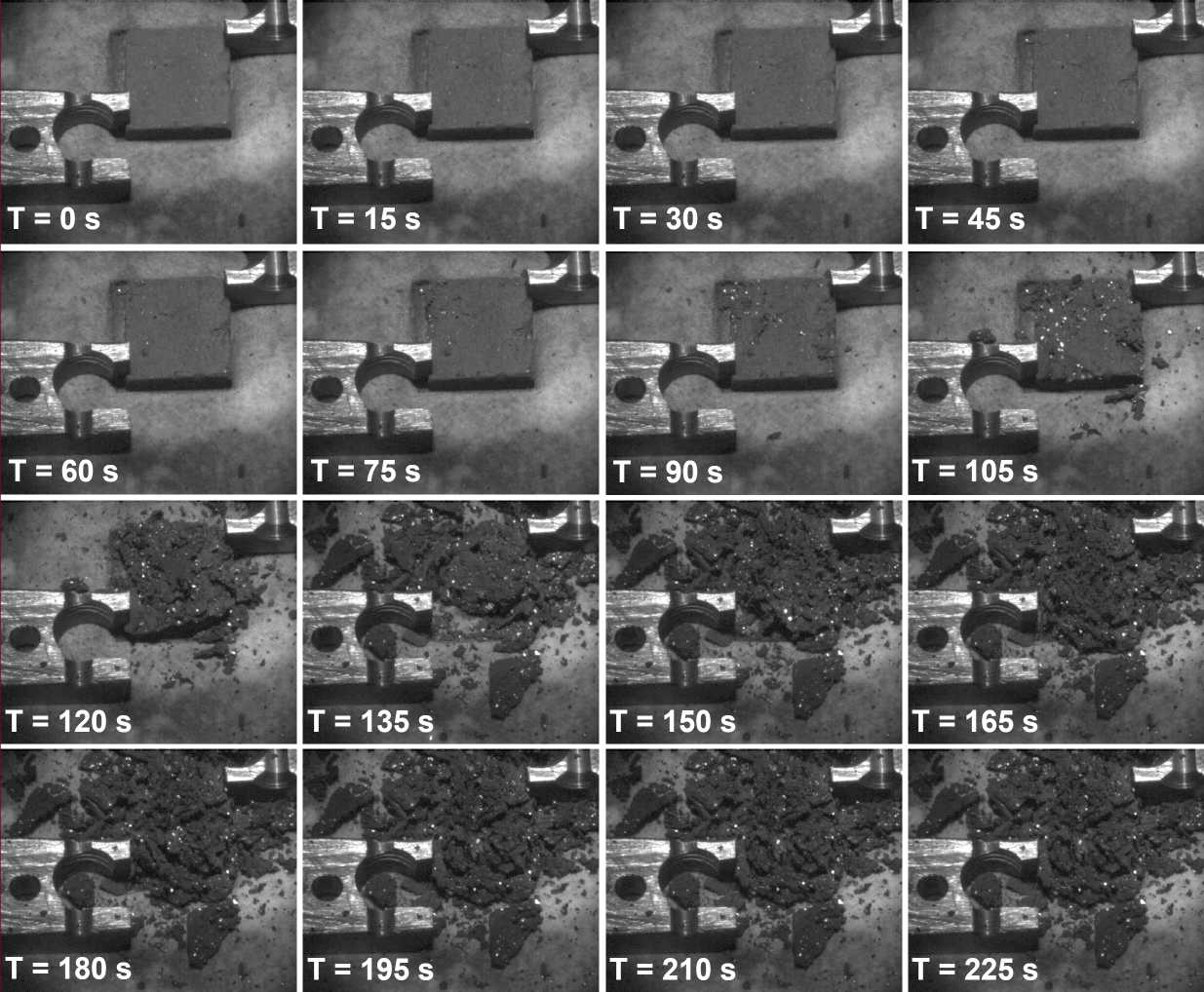}
\caption{Snapshots of an experiment with CI simulant and an irradiance equivalent to that of the Sun at a heliocentric distance $r=0.14\au$. \label{fig:snapshots}}
\end{figure}

The destruction process can be divided into three phases based on a qualitative analysis of the video recordings \citep{video}. The first phase starts from the moment the irradiance begins. It mostly involves the initial heating of the samples, with a few minuscule ejections of presumably loosely bound dust particles from the surface. In Fig.~\ref{fig:snapshots}, this phase occurs between $t=0\second$ and $t=90\second$. 

The second phase begins when the surface temperature of the sample has passed a threshold at which the rate of disruption increases drastically. Larger fragments, of the order of a few mm in size, are ejected in multiple explosive events between $t=105\second$ and $t=180\second$ (Fig.~\ref{fig:snapshots}). Although we have not quantified the ejection-velocity distributions of the particles during these explosive events, it is evident from the video footage, and the spreading of fragments in the vacuum chamber, that a significant fraction of them have high enough velocities to clear the irradiated area \citep{video}. During this second phase the frequency of such ejection events is higher than in the first phase as fresh material is continuously exposed to the incident light. The frequency of ejections then starts decreasing, either because all of the sample material has been eroded away and has been expelled completely from the test area, or because fragments from previous events have accumulated on top of the remaining sample, effectively shadowing it from the incident light. In either case, the rate of ejections seems to approximately follow a normal distribution. It is noteworthy that fragments that have been ejected in previous events, but remain in the irradiated area, may experience some secondary disruption events, but they do not completely disintegrate; their sizes are of the order of millimeters, which is much larger than the particle size of the initial unmixed simulant powder. Instead it appears that these fragments become inert and are unaffected by subsequent exposure to the incident light. 

The third phase is what follows when eroded, inert material has remained on top of the yet unaffected sample. The fresh material cannot be directly heated by the light, but the continuous influx of radiative power appears to gradually propagate through the inert layers. The process slowly raises the temperature of the sub-surface, leading to the fragmentation of the entire sample. We do not observe energetic and explosive sub-surface events that would eject the overlaying material, but rather small movements and overall expansion of the pile of material.

\subsection{Quantitative assessment of the destruction time}

\begin{figure}
\centering
\includegraphics[width=\textwidth]{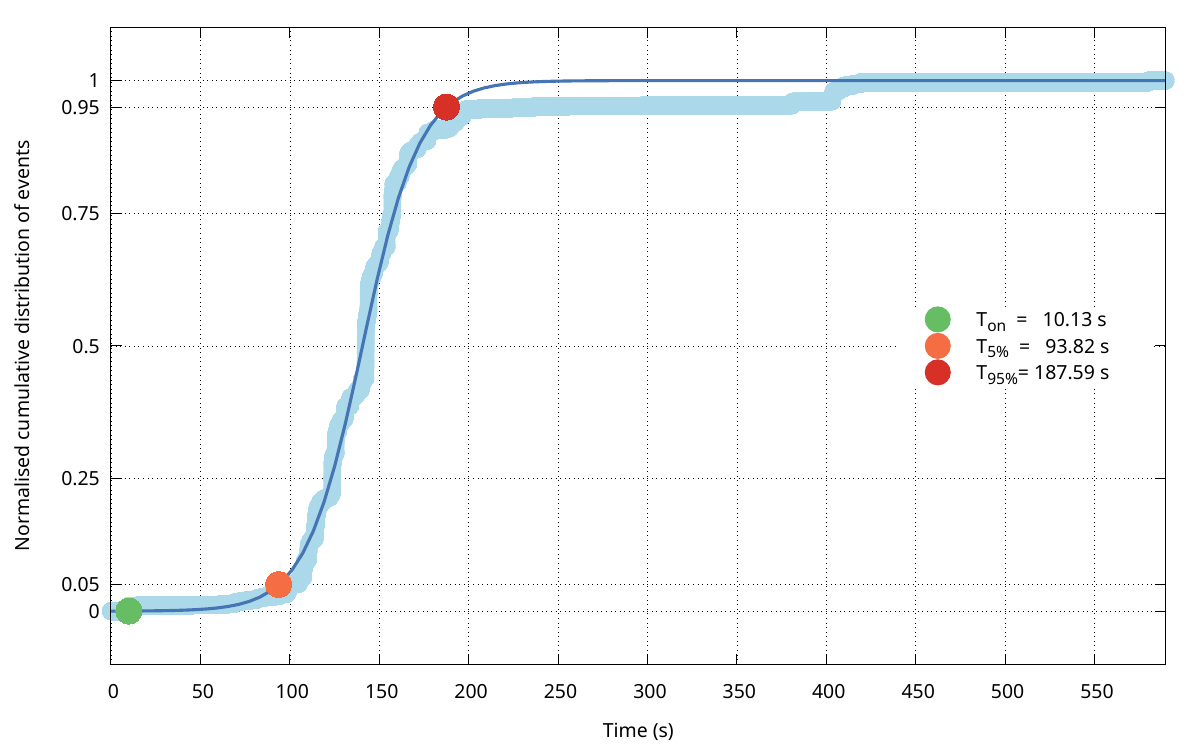}
\caption{An example of the measured activity as a function of time at an irradiance level corresponding to $r=0.14\au$. The normalized cumulative distribution of the detected events (cyan) and the fitted sigmoid function (blue). The colored dots represent the beginning of each of the three phases of the disruption process as described. \label{fig:cumulexp}}
\end{figure}

A key objective of our experiments is to accurately measure the destruction time at different irradiance levels. To achieve the objective, we analyzed the video footage recorded in real time at 60 frames per second, and determined the time it takes for each sample to be destroyed---measured from the moment the irradiance begins until the activity ceases. To detect the activity we first stabilize the video files to remove potential shaking from the cameras, and then we make use of the motion-detection algorithm in the open-source multimedia player \verb|VLC|. The algorithm detects moving shapes between each pair of subsequent frames in the video, and outputs the number of such moving shapes in each frame. From the time series of the number of moving shapes in each video frame, which approximately follows a normal distribution, we then derive the cumulative distribution (Fig.~\ref{fig:cumulexp}). This is done to smooth the noise in the normal distribution which is generated by frames with no moving shapes, as the activity is not constant. The cumulative distribution is then fitted with a sigmoid function of the general form
\begin{equation}
f(t)=a+\frac{b}{1+\exp{((d-t)/c)}}\,,    
\end{equation}
where $a$, $b$, $c$, and $d$ are the fitting parameters. In our application, $a$ is pre-set to zero and $b$ is pre-set to the maximum value of the cumulative distribution, as these are the boundary conditions, leaving parameters $c$ and $d$ to fit the data. After fitting, we assign the times at which the aforementioned phases of the disruption process happen as follows. The first phase begins when the sample irradiation begins. The second phase starts at the time when the fitted sigmoid function reaches 5\% of its maximum value, determined using the inverse of the fitted sigmoid function which was derived analytically and with the same fitting parameters. In the same manner we assign the time at which the sigmoid function reaches 95\% of its maximum value to the end of phase three, which we consider the end of the disruption process. In this way, phases two and three represent together the central 90\% of the activity duration. Although the choice to use the 5th and the 95th percentiles is arbitrary, it has proven to be a reasonably good proxy for measuring the destruction time. In addition, we are mostly interested in relative comparisons between the different experiment runs, and they are not being affected by the choice of percentiles. As we have video footage from two cameras for each experiment, we perform the same analysis for both videos. The derived destruction time from the two orthogonal viewing angles are nearly identical for each experiment, verifying that the motion detection algorithm is robust enough for our purpose. We nevertheless assign to each experiment the average destruction time as derived from the two videos analyzed separately. An example of the process is shown in Fig.~\ref{fig:cumulexp} for the same experiment that is shown in Fig.~\ref{fig:snapshots} at $r=0.14\au$.

\subsection{Destruction time as a function of heliocentric distance}

\begin{figure}
\centering
\includegraphics[width=\textwidth]{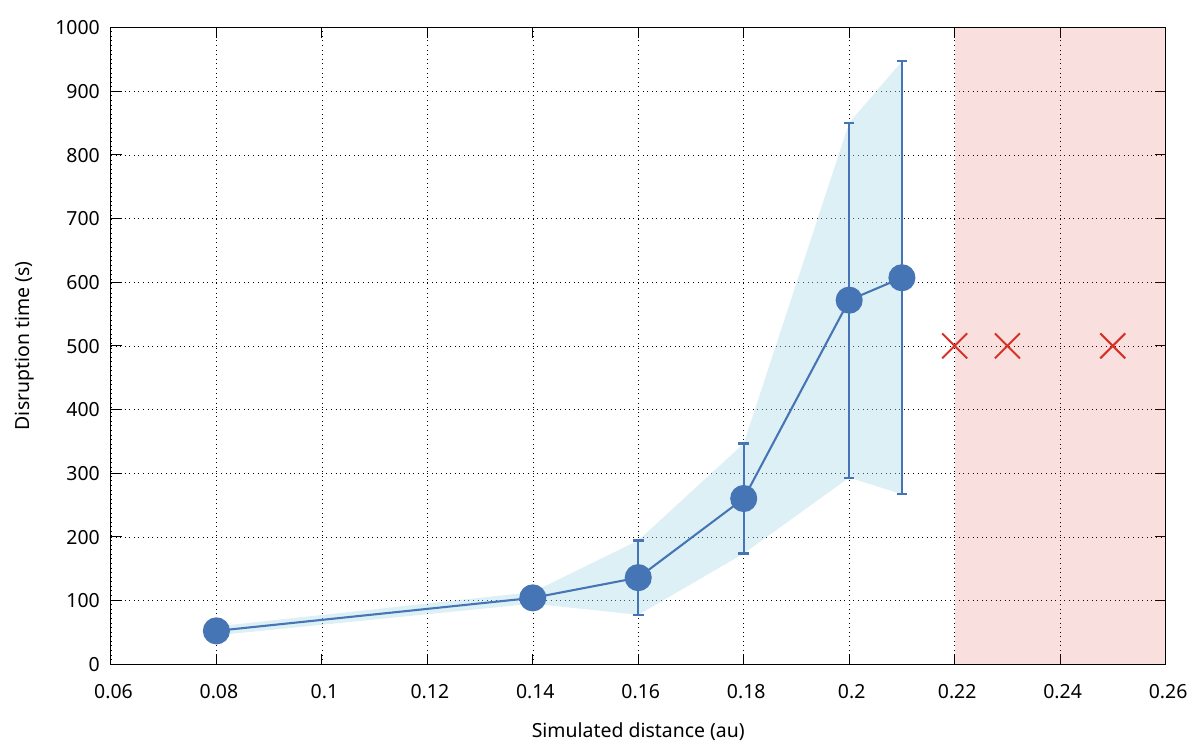}
\caption{Disruption time versus simulated heliocentric distance. The red crosses do not provide information about the length of the experiments.}\label{fig:time-distance}
\end{figure}

We then proceeded to carry out experiments at different simulated heliocentric distances in the range $0.07\au < r < 0.25\au$. The experiments allow us to establish a correlation between the destruction time and the heliocentric distance. The average destruction time for 3--4 experiments at each simulated heliocentric distance is shown in Fig.~\ref{fig:time-distance}. As expected, the thermal erosion of the CI simulant proceeds at a faster rate at smaller heliocentric distances. Moreover, we observe that the disruption process is more energetic altogether at $r<0.1\au$, with nearly all of the sample material being directly ejected out of the light spot. The aforementioned third phase is thus entirely missing. On the other end of the range of simulated heliocentric distances, the destruction time is significantly longer. The erosion is non-continous and happens in well-separated events. There is no second phase as described above, but only a long third phase. At the very end of our range in heliocentric distance, $r>0.22\au$, the sample remains intact even after three hours of irradiation.

\subsection{Geochemical and mineralogical analysis}

\begin{figure}
\centering
\includegraphics[width=\textwidth]{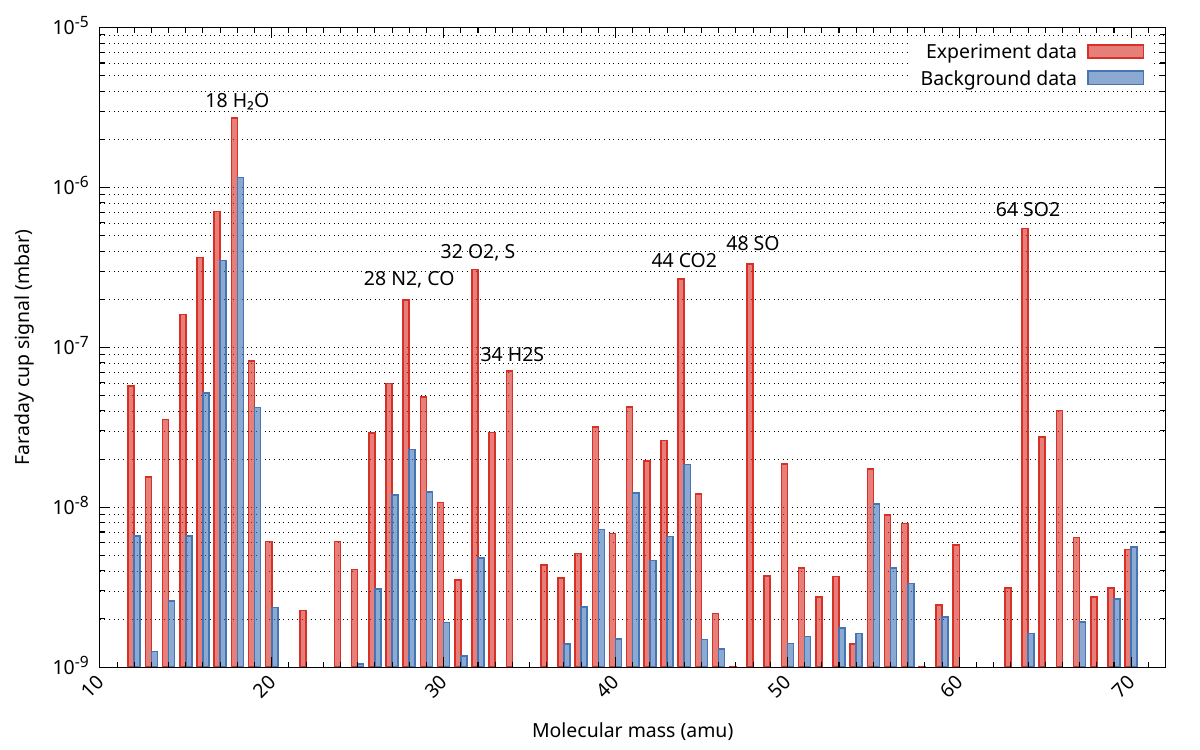}
\caption{Residual gas analysis scanned peak intensities immediately before (blue) and during irradiation (red). The compounds corresponding to some of the highest intensities are labeled.   \label{fig:rga}}
\end{figure}

During the destruction process of the samples presented above, the pressure inside the chamber increases from $P<5\cdot10^{-5}\,$mbar to $P\approx5\cdot10^{-3}\,$mbar due to outgassing of volatiles from the samples. Our vacuum chamber is equipped with a residual gas analyzer (HIDEN Analytical HAL-200RC), which can be used to study the composition of the released volatiles. However, since it can only operate at pressures lower than $P\approx10^{-4}\,$mbar, we could not take direct measurements during the experiments we presented above. To circumvent this limitation we performed additional experiments with the same simulant but with much smaller pieces $(D\approx3\mm)$, and measured the composition of the chamber atmosphere before and during irradiation. Fig.\ref{fig:rga} shows an example of the raw data as they are output by the RGA control software, to provide a qualitative assessment of the outgassing. The peaks with the largest detected increase in intensity correspond to species such as H$_2$O, CO$_2$, CO, sulfur compounds, and possibly some organics. These species are present due to the thermal decomposition of serpentine, sulfides, and coal, in agreement with the thermogravimetric measurements by \citet{Britt2019} and the models of \citet{MacLennan2024}.

Some noticeable changes were observed upon inspection of the samples: the pyrite grains are much more visible on the fractured flakes (Fig.~\ref{fig:snapshots}). Unlike the fresh CI simulant, which is highly dispersible in water, the irradiated simulant is completely indispersible, and does not discolor the water when submerged.

\section{Discussion}

Our experiments show that asteroids composed of CI-type carbonaceous chondrites undergo rapid erosion at small heliocentric distances, solely driven by electromagnetic radiation. Furthermore, thermally-driven erosion can be virtually immediate provided that the irradiance is high enough. That is, the process does not require the material to be previously weakened by more gentle processes such as micrometeoroid bombardment \citep{Wiegert2020} or thermal fatigue due to temperature variations over many rotational periods or even orbital periods \citep{Delbo2014}.

We can extrapolate the disruption-time measurements obtained in the laboratory to gain some insight into the efficiency of this irradiance-driven disruption mechanism in eroding entire asteroids. Since (3200)~Phaethon has been shown to exhibit activity for about three days during each of its perihelion passages \citep{Li2013}, we can use its average physical diameter of $5.4\pm0.2\km$ \citep{MacLennan2022} and perihelion distance of 0.14~au as a reference case for comparison. Considering a spherical body with half of its area illuminated we calculated the areas of the spherical zones that receive irradiances corresponding to the distances we used in our experiments, from 0.14~au at the sub-solar point to 0.21~au at a polar angle of $63.6^\circ$ with respect to the Sun vector. We then calculated the average erosion time for a hypothetical sample pellet on the asteroid surface by taking the harmonic mean of the erosion times given in Fig.~\ref{fig:time-distance} weighted by the fractional area of the corresponding zones, yielding $\bar{t}_{0.14}=166\second$. Combined with the mass of the sample pellets of $3.5\g$ and their area of $4\cm^2$ the erosion rate would be approximately $50\g \meter^{-2} \second^{-1}$ or $1.3\times10^6\kg\second^{-1}$ for the entire asteroid. This value is much larger than the observed mass-loss rate of (3200)~Phaethon of  $3\kg\second^{-1}$ \citep{Jewitt2013}. The large overestimation of the mass-loss rate by our experiments can be attributed to two key factors. First, in our experiments we use fresh CI asteroid simulant material, whereas Phaethon's surface has been shown to resemble CY chondrites \citep{MacLennan2024} that show evidence for thermal alteration \citep{King2019}, and may thus not break up as easily as CI chondrites when irradiated. Second, in our experiments we noticed that the destruction of the sample does not necessarily lead to an ejection of all the fragments from the surface, as most of them remain within the irradiated area (Fig.~\ref{fig:snapshots}). 

We can make a similar comparison of our experimental results to the case of 322P/SOHO~1 (P/1999~R1), which has a perihelion distance of $0.05\au$. Assuming it is a dark asteroid with a CI chondrite-like composition, and thus adopting a diameter of $320\meter$ \citep{knight2016}, our experimental results would predict an erosion rate of $28\times10^3\kg\second^{-1}$. The predicted erosion rate is only one order of magnitude larger than the measured mass-loss rate of $\leq2\times10^3\kg\second^{-1}$ \citep{knight2016}, suggesting that a direct irradiance-driven disruption similar to what we observe in our experiments could be responsible for the brightening of 322P/SOHO~1 during its perihelion passage. We note that the proposed mechanism could also explain why activity could not be detected at heliocentric distances ranging from $1.2\au$ to $2.2\au$ \citep{knight2016}.

\begin{figure}
\centering
\includegraphics[width=\textwidth]{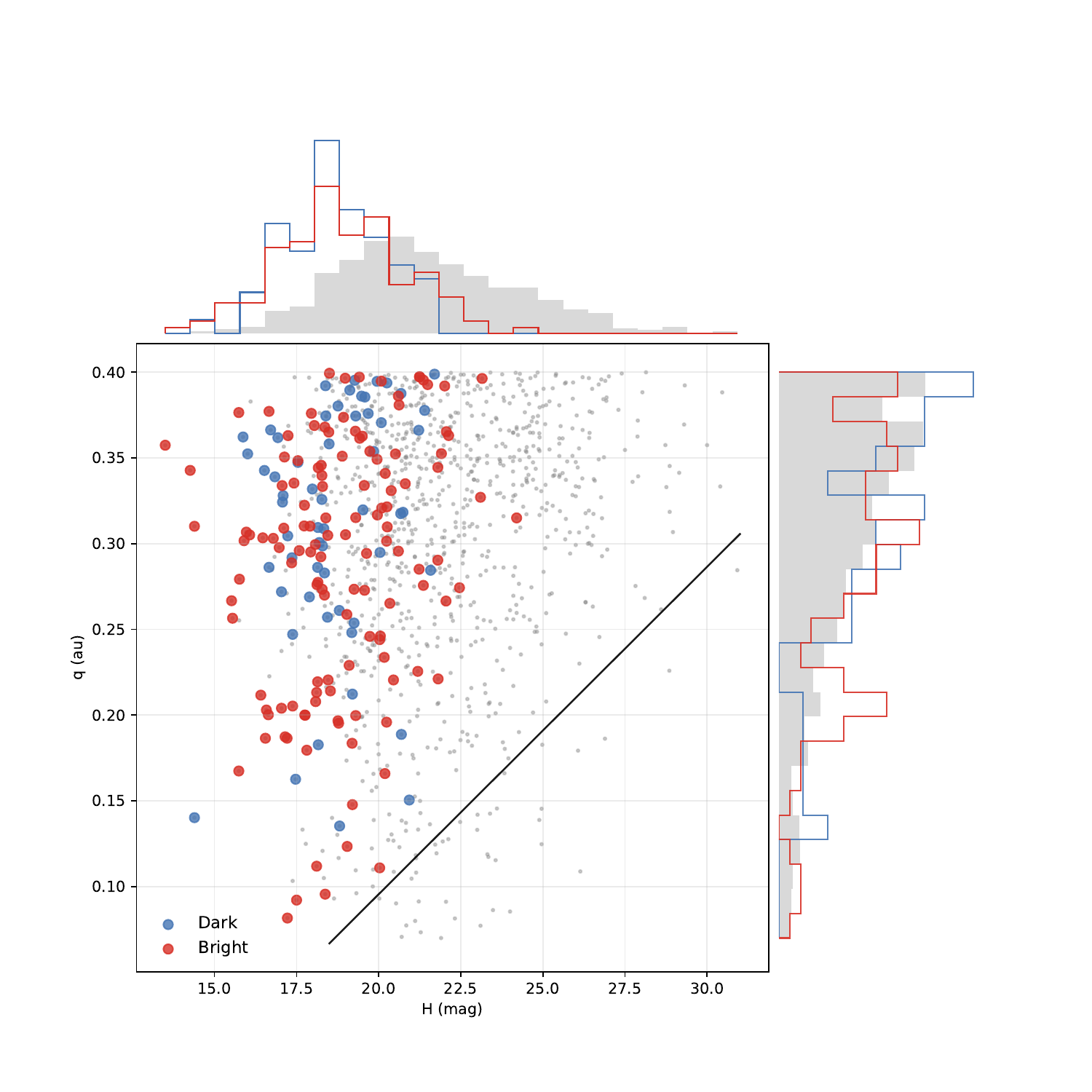}
\caption{The perihelion distances, $q$, and absolute magnitudes, $H$, for all near-Earth asteroids (NEAs) with $q<0.4\au$ are shown with gray dots. NEAs with spectral types C, B, D, P, K, and L and/or with albedos $p_v<0.1$ are assigned as dark and are highlighted with blue points and NEAs with spectral types S, Q, V, R, and M and/or with albedo $p_v>0.1$ are assigned as bright and are highlighted with red points. The black line corresponds to the average disruption distance as a function of $H$ \citep{Granvik2016}. The histograms are showing the distributions of the respective groups of asteroids in absolute magnitude and perihelion distance, and are scaled by area.   \label{fig:H_q}}
\end{figure}

A rapid erosion rate suggests that there should not exist many carbonaceous asteroids with small heliocentric distances, $q$. The prediction can be tested by examining the orbital and physical parameters of near-Earth asteroids from the Minor Planet Center and SsODNet \citep{2023A&A...671A.151B} databases. We selected all numbered and unnumbered asteroids with $q<0.4\au$, resulting in 1103 asteroids. Out of these, we were able to extract information about the geometric albedos and/or spectral classifications for 202 asteroids: geometric albedos for 133 asteroids and spectral classifications for 129 asteroids (Fig.~\ref{fig:H_q}). Both geometric albedo and spectral classification is available for 61 of the 202 asteroids. 

There is an apparent dearth of dark objects and objects belonging to the C-complex at very small perihelion distances, which is in line with a rapid erosion of such bodies close to the Sun. However, there is also a dynamical preference for bright asteroids to reach smaller perihelion distances than dark asteroids \citep{toliou2023resonant}. A larger sample of NSA spectral classifications and albedos are needed to assess if the scarcity of dark objects is caused by dynamical or thermal effects.

\section{Conclusions}

The surfaces of asteroids composed of CI-like material can be eroded solely by the intense electromagnetic radiation by the Sun when they reach sufficiently small heliocentric distances, $r$. Immediate erosion, that is, not accounting for effects arising from cyclic temperature changes, occurs at $r\lesssim0.2\au$. Given the relatively high rate of erosion, the same mechanism can eventually destroy entire asteroids. The super-catastrophic disruption of asteroids at small perihelion distances \citep{Granvik2016} could thus potentially be explained by purely thermally-driven mechanisms without the need to resort to other explanations such as meteoroid bombardment \citep{Wiegert2020}. 

While our results show that a thermally-driven mechanism is a viable explanation for activity on NSAs and even for their complete destruction, there remain a number of open questions that need to be resolved in the future such as how the measured disruption rate varies with composition and texture (monolithic versus regolith-like). We also need to measure the ejection speeds and sizes of the fragments ejected in our experiments to more accurately predict the mass-loss rate from entire asteroids. 

To test the hypothesis that the brightening of 322P during its perihelion passage is due to instantaneous thermally-driven erosion, we need to establish its taxonomic type with more certainty through, e.g., reflectance spectroscopy in the visual to near-infrared wavelengths---possible in July 2027---and model the erosion more accurately by including, for example, information about the fragment ejection speeds and sizes. 

\section{Acknowledgements}
The authors would like to thank two anonymous reviewers for their constructive comments that helped improve this manuscript. 

The work of MG, LS and AH was supported by the Swedish Research Council under grant number 2022-04615. The work of JG was supported by the  Carl Trygger Foundation for Scientific Research under grant number CTS 24:3838.

\end{document}